\documentstyle[12pt,prd,aps,epsfig,preprint]{revtex}

\begin{document}
\draft
\def\ds{\displaystyle}
\date{\today}
\title{\bf Exact solution of the time-dependent harmonic plus an
inverse harmonic potential with a time-dependent electromagnetic field}
\author{Cem Y\"{u}ce\thanks{e-mail: yuce@metu.edu.tr}}
\address{Physics Department, Middle East Technical University,
06531 Ankara, Turkey}

\maketitle
\begin{abstract}
In this article, the problem of the charged harmonic plus an
inverse harmonic oscillator with time-dependent mass and frequency
in a time-dependent electromagnetic field is investigated. It is
reduced to the problem of the inverse harmonic oscillator with
time-independent parameters and the exact wave function is
obtained.
\end{abstract}

\thispagestyle{empty} ~~~~\\ 
\narrowtext
\newpage
\setcounter{page}{1}

\section{Introduction}
A time-dependent quantum system with a Hamiltonian which contains a singular term
\begin{eqnarray}\label{e1}
\hat{H}=\frac{\hat{\textbf{p}}^2}{2m(t)}+ \frac{m(t)}{2}
\omega^2(t)~ \hat{\textbf{q}}^2+\frac{C}{m(t) \hat{\textbf{q}}^2},
\end{eqnarray}
is one of the rare examples admitting exact solution and has been subject of great
interest in a variety of contexts
\cite{R4801,R4802,R4803,R4804,R4805,R4806,R4807,R4808,R4809,R4810}, because of its
various applications in different areas of physics. For example, it is used for
constructing exactly solvable models of interacting many-body systems
\cite{R4801,R4802}, for modelling polyatomic-molecules \cite{R4803} and for
coherent-state formalism \cite{R4811}. The Hamiltonian in Eq. (1) has been studied and
the exact solution was obtained by using the Lewis and Riesenfeld invariant method and
by making canonical and unitary transformations \cite{R4805,R4806,R4807,R4808,R4809}.
Although the exact solution of the time-dependent charged harmonic oscillator in a
time-varying electromagnetic field is known \cite{R4812}, the charged harmonic plus
inverse harmonic oscillator problem in a time-varying electromagnetic field has not
yet been solved exactly.

In this letter we take the following Hamiltonian into account
\begin{eqnarray}\label{e2}
\hat{H}=\frac{\left[\hat{\textbf{p}}-q\hat{\textbf{A}}(t)\right]^2}
{2m(t)}+\frac{m(t)}{2} \omega^2(t)~ \hat{\bf{q}}^2+\frac{C}{m(t)
\hat{\textbf{q}}^2}+q \hat{\phi}(t)=i\frac{\partial}{\partial t},
\end{eqnarray}
where we have set $\hbar=1=c$. By making some transformations,
Eq. (2) is reduced to
\begin{eqnarray}\label{e3}
\frac{\hat{\bf{p}^{\prime}}^2}{2}+\frac{C}{\hat{\bf{q}}^{\prime 2}}
=if (t^{\prime}) \frac{\partial}{\partial t^{\prime}},
\end{eqnarray}
where $f(t)$ is a time-dependent function. Finally, Eq. (3) is
integrated.

\section{Method}
The Hamiltonian for the charged harmonic plus inverse harmonic
oscillator constrained to move in the $x-y$ plane is given
\begin{eqnarray}\label{e4}
\hat{H}=\frac{\left[\hat{\textbf{p}}-q\hat{\textbf{A}}(t)\right]^2}
{2m(t)}+\frac{m(t)}{2} \omega^2(t)~(\hat{x}^2+\hat{y}^2)+
\frac{C}{m(t)(\hat{x^2}+\hat{y^2})}+q\hat{\phi}(t)
=i\frac{\partial}{\partial t},
\end{eqnarray}
where $C$ is any constant and~ ${\ds\hat{\textbf{p}}=
-i\frac{\partial}{\partial x}\hat{\textbf{e}_1}-
i\frac{\partial}{\partial y} \hat{\textbf{e}_2}}$.

In this letter, we restrict ourselves to the case where the
electromagnetic potentials are given by
\begin{eqnarray}\label{e5}
\hat{\textbf{A}}=\frac{B(t)}{2} (\hat{y}{\hat{\textbf{e}}}_1
-\hat{x}{\hat{\textbf{e}}}_2), ~~~~~\hat{\phi}(t)=0,
\end{eqnarray}
where~ $(\hat{\textbf{e}}_1,\hat{\textbf{e}}_2,\hat{\textbf{e}}_3)$
are the unit vectors of the cartesian basis. With this choice, the
magnetic and electric fields turn out to be
\begin{equation}
{\bf B}= B(t)\hat{\textbf{e}_3},
\end{equation}
\begin{equation}
{\bf E}=\frac{1}{2} \rho \frac{d B(t)}{dt} {\bf \hat{\phi}},
\end{equation}
where $({\bf \hat{\rho}},{\bf \hat{\phi}},{\bf \hat{e}_3)}$ form
the cylindrical basis. Hereafter, we will denote $\omega(t)$, $m(t)$
and $B(t)$ by $\omega$, $m$ and $B$, respectively, for brevity. By
substituting Eq. (5) into Eq. (4), the Hamiltonian becomes
\begin{eqnarray}\label{e7}
\frac{\hat{p_x}^2+\hat{p_y}^2}{2m}-\frac{q B}{4 m}~ (\hat{y}
\hat{p_x}-\hat{x} \hat{p_y})+\frac{m}{2}\left(\omega^2+
\frac{q^2 B^2}{4m^2}\right)(\hat{x}^2+\hat{y}^2)+\frac{C}{m
(\hat{x}^2+\hat{y}^2)}=i\frac{\partial}{\partial t}
\end{eqnarray}
Now we want to make a transformation to get rid of the cross-term in Eq. (8). To do
this, an orthogonal transformation is introduced \cite{R4812}.
\begin{eqnarray}\label{e8}
\pmatrix{ \hat{x}   \cr
                     \hat{y}   \cr}=
\pmatrix{ \cos \beta(t) & -\sin \beta(t)  \cr
                     \sin \beta(t) & \cos \beta(t)   \cr}
\pmatrix{\hat{x^{\prime}}  \cr
                       \hat{y^{\prime}}   \cr},
\end{eqnarray}
The time-dependent transformation (9) implies the following
transformations for the quadratic terms
\begin{eqnarray}\label{e9}
\hat{p_x}^2+\hat{p_y}^2 \rightarrow \hat{p_{x^{\prime}}}^2
+\hat{p_{y^{\prime}}}^2,~~~~
\hat{x}^2+\hat{y}^2 \rightarrow \hat{x}^{\prime^2}+\hat{y}^{\prime 2},
\end{eqnarray}
and for the cross-term and time derivative
\begin{equation}
\hat{y} \hat{p_{x}}-\hat{x} \hat{p_{y}} \rightarrow \hat{y}^{\prime}
\hat{p}_{x^{\prime}}-\hat{x}^{\prime} \hat{p}_{y^{\prime}}
\end{equation}
\begin{equation}
\frac{\partial}{\partial t} \rightarrow \frac{\partial}{\partial t}+
i\dot{\beta(t)}\left[\hat{y}^{\prime}\hat{p}_{x^{\prime}}-
\hat{x}^{\prime}\hat{p}_{y^{\prime}}\right].
\end{equation}
To eliminate the cross-term from the Eq. (8), $\beta$ is chosen as
\begin{eqnarray}
\label{e11}\dot{\beta}(t)=\frac{q B}{4m},
\end{eqnarray}
where dot denotes time derivation as usual. In this case, Eq. (8)
becomes
\begin{eqnarray}\label{e12}
\frac{\hat{p_{x^{\prime}}}^2+\hat{p_{y^{\prime}}}^2}{2m}+
\frac{m}{2}\left(\omega^2+\frac{q^2 B^2}{4m^2}\right)
(\hat{x}^{\prime 2}+\hat{y}^{\prime 2})+\frac{C}
{m(\hat{x}^{\prime 2}+\hat{y}^{\prime 2})}=i\frac{\partial}{\partial t}.
\end{eqnarray}
Eq. (14) does not depend on the cross-term and the square of the angular frequency is
shifted by an amount of~ ${\ds q^2B^2/(4m^2)}$.\\
To remove the harmonic part, we make the transformation
\begin{equation}
\hat{p_x}^{\prime}\rightarrow \hat{P_1} = \hat{p_x}^{\prime}+
i\alpha(t)\hat{x}^{\prime},~~~\hat{p_y}^{\prime}\rightarrow \hat{P_2}
=\hat{p_y}^{\prime}+i\alpha(t) \hat{y}^{\prime},
\end{equation}
\begin{equation}
\hat{x}^{\prime} \rightarrow  \hat{Q_1} =
\hat{x}^{\prime},~~~~~~~~~~~~~\hat{y}^{\prime}
\rightarrow \hat{Q_2}~ = \hat{y}^{\prime},
\end{equation}
where $\alpha(t)$ is a time-dependent function to be determined later. Since ${\ds
\hat{\textbf{p}}=-i\frac{\partial}{\partial x}
\hat{\textbf{e}_1}-i\frac{\partial}{\partial y}\hat{\textbf{e}_2}}$, we observe that
the transformation (15) is equivalent to the multiplication of the wave function by
~${\ds\exp\left[-\alpha(t)(x^{\prime 2}+y^{\prime 2})/2\right]}$.\\
Under the transformation (15), the time derivative operator becomes
\begin{eqnarray}\label{e14}
\frac{\partial}{\partial t} \rightarrow \frac{\partial}{\partial t}
-\dot{\alpha}(t)\left(\frac{\hat{Q_1}^{\prime 2}+\hat{Q_2}^{\prime 2}} {2}\right),
\end{eqnarray}
Substituting Eqs. (15), (16) and (17) in (14), we get
\begin{eqnarray}\label{e15}
\frac{\hat{P_1}^2+\hat{P_2}^2}{2m}+ \Omega^2(t)\left[\hat{Q_1}^2+
\hat{Q_2}^2\right]+\frac{C}{m(\hat{Q_1}^2+\hat{Q_2}^2)}+
i\frac{\alpha(t)}{m}(\hat{Q_1}\hat{P_{1}}+\hat{Q_2}
\hat{P_{2}})=i\frac{\partial}{\partial t}+\frac{\alpha(t)}{m},
\end{eqnarray}
where
\begin{eqnarray}
\nonumber \Omega^2(t)=\frac{m}{2}\left(\omega^2+\frac{q^2 B^2}{4m^2}\right)
-\frac{\alpha^2(t)}{2m}+i\frac{\dot{\alpha}(t)}{2}.
\end{eqnarray}
The time-dependent functions $\alpha(t)$ should be chosen so that
harmonic term in the Hamiltonian vanishes.
\begin{eqnarray}\label{e16}
\frac{m}{2}(\omega^2+\frac{q^2 B^2}{4m^2})-
\frac{\alpha^2(t)}{2m}+i\frac{\dot{\alpha(t)}}{2}=0.
\end{eqnarray}
Eq. (18) is simplified after the elimination of the harmonic part.
\begin{eqnarray}\label{e17}
\frac{\hat{P_1}^2+\hat{P_2}^2}{2m}+\frac{C}{m\left(\hat{Q_1}^2+
\hat{Q_2}^2\right)}+i\alpha(t)\left[\hat{Q_1}\hat{P_{1}}+
\hat{Q_2}\hat{P_{2}}\right]=i\frac{\partial} {\partial t}+
\frac{\alpha(t)}{m}.
\end{eqnarray}
Now a velocity-dependent interaction term is contained in Eq. (20).
To get rid of this term, we need another transformation.
\begin{equation}
\hat{P_1} \rightarrow \hat{P_1}^{\prime}=\mu(t) \hat{P_1}, ~~~~~~~\hat{P_2}
\rightarrow \hat{P_2}^{\prime}=\mu(t)\hat{P_2}
\end{equation}
\begin{equation}
\hat{Q_1} \rightarrow \hat{Q_1}^{\prime}=\frac{\hat{Q_1}}{\mu (t)} , ~~~~ \hat{Q_2}
\rightarrow \hat{Q_2}^{\prime}=\frac{\hat{Q_2}} {\mu(t)},
\end{equation}
then the time derivative operator becomes
\begin{eqnarray}\label{e19}
\frac{\partial}{\partial t} \rightarrow \frac{\partial} {\partial
t}-i\frac{\dot{\mu}(t)}{\mu(t)} (\hat{Q}_1^{\prime}
\hat{P}_1^{\prime}+\hat{Q}_2^{\prime}\hat{P}_2^{\prime}).
\end{eqnarray}
If $\mu(t)$ is chosen so that
\begin{eqnarray}\label{e20}
\alpha(t)=-\frac{\dot{\mu}(t)}{\mu(t)},
\end{eqnarray}
then the velocity dependent interaction term vanishes. The new
equation is just the equation of a particle under an inverse
harmonic potential and with a time-dependent mass.
\begin{eqnarray}\label{e21}
\hat{P_1}^{\prime 2}+\hat{P_2}^{\prime 2}+
\frac{2C}{\hat{Q_1}^{\prime 2}+\hat{Q_2}^{\prime 2}}=i
2m \mu^2(t) \frac{\partial}{\partial t}+2\mu^2(t) \alpha(t)=k^2.
\end{eqnarray}
Since left-hand side and right-hand side of (25) depend on
different parameters, the two sides are equal to a constant,
$k^2$. We can rewrite Eq. (25) as a wave equation rather than an
operator equation
\begin{eqnarray}\label{e22}
-\frac{\partial^2 \Psi(x,y,t)}{\partial Q_1^{\prime 2}}-\frac{\partial^2
\Psi(x,y,t)}{\partial Q_2^{\prime 2}}
+\frac{2C}{Q_1^{\prime 2}+Q_2^{\prime 2}}\Psi(x,y,t)=i 2m \mu^2
\frac{\partial \Psi(x,y,t) }{\partial t}+2\mu^2 \alpha \Psi(x,y,t).
\end{eqnarray}
Using the separation of variables technique, we can solve Eq. (26).

Let us write $\Psi=R(Q_1^{\prime},Q_2^{\prime})T(t)$
\begin{eqnarray}\label{e23}
T(t)=\exp\left[- i\int_0^{t}\frac{k^2+2\mu^2(t^{\prime})\alpha(t^{\prime})}
{2m(t^{\prime}){\mu^2(t^{\prime})}} dt^{\prime}\right].
\end{eqnarray}
Coordinate-dependent  part of Eq. (26) can be decomposed in polar
coordinates
\begin{equation}
Q^2 \frac{d^2 Z}{dQ^2}+Q \frac{d Z}{d Q}+(k^2 Q^2-2C-n^2) Z=0
\end{equation}
\begin{equation}
\frac{d^2 \Theta}{d\theta^2} = -n^2 \Theta,
\end{equation}
where $Q^2=Q_1^{\prime 2}+Q_2^{\prime 2}$,~
$Q_1^{\prime}=Q\cos\theta$,~$Q_2^{\prime}=Q \sin \theta$ and
$R(Q_1^{\prime},Q_2^{\prime})=Z(Q)\Theta(\theta)$.

The relation between $\theta$ and $x,y$ is given
\begin{eqnarray}\label{e25}
\tan \theta=\frac{Q_2^{\prime}}{Q_1^{\prime}}=
\frac{\cos\beta x+\sin\beta y}{-\sin\beta x+\cos \beta y}
\end{eqnarray}
where $\beta$ was defined in Eq. (13).

If we introduce $\nu^2=2C+n^2$, we see that Eq. (28) is just the Bessel differential
equation of order $\nu$ \cite{R4813}. The solutions of (28,29)
\begin{eqnarray}\label{e26}
Z(Q)=A J_\nu(k Q)+B N_\nu(k Q)\\
\Theta=C \exp (\pm i n \theta),
\end{eqnarray}
where $J_\nu(k Q^{\prime})$, $N_\nu(k Q^{\prime})$ are  the Bessel
function of the first and the second kind respectively. The
constants $A$, $B$ and $C$ are determined from the boundary
conditions. Note that $\nu$ and $n$ are not the same. If the
inverse harmonic potential term goes to zero, they become equal to
each other.

Finally, the complete exact solution of Eq. (4) is given
\begin{eqnarray}\label{e27}
\Psi=\left[A J_\nu\left(k\frac{\sqrt{x^2+y^2}}{\mu(t)}\right)+
B N_\nu\left(k\frac{\sqrt{x^2+y^2}}{\mu(t)}\right)\right]
\exp[\alpha(t)(x^2+y^2) \pm i n\theta-if]
\end{eqnarray}
where ${\ds f(t)=\int_0^t\frac{k^2+2 \mu^2(t^{\prime})\alpha(t^{\prime})}
{2m(t^{\prime})\mu^2(t^{\prime})} dt^{\prime}}$ and $\alpha$, $\mu$,
$\theta$ were defined Eqs. (19), (24) and (30) respectively.

In summary, the exact solution of the problem of the charged
harmonic plus an inverse harmonic oscillator with time-dependent
mass and frequency in a time-dependent electromagnetic field is
obtained by making some appropriate transformation on momentum and
coordinate operators.

\end{document}